# The Effect of ICT Self-Discipline in the Workplace

**Balsam Al-Dabbagh**
School of Information Management
Victoria University of New Zealand
Wellington, New Zealand
Email: balsam.aldabbagh@vuw.ac.nz

**Eusebio Scornavacca**
Merrick School of Business
University of Baltimore
Baltimore, USA
Email: escornavacca@ubalt.edu

**Allan Sylvester**
School of Information Management
Victoria University of New Zealand
Wellington, New Zealand
Email: allan.sylvester@vuw.ac.nz

**David Johnstone**
School of Information Management
Victoria University of New Zealand
Wellington, New Zealand
Email: david.johnstone@vuw.ac.nz

## Abstract

The ubiquity of Information and Communication Technologies (ICTs) in work settings has changed the way employees behave. ICTs such as smartphones, email and social media makes employees more connected than has ever been possible. The degree of ICT connectivity can create a positive as well as a negative impact on employees' productivity.  In this context, the notion of ICT self-discipline – an individual's ability to regulate their behaviours towards ICTs – becomes pivotal in the process of managing ICT connectivity. This follow-on study reports the results of an online survey of 443 to New Zealand professionals regarding the influence of ICT self-discipline on the relationship between ICT connectivity and employee productivity. Findings indicated that the impact of ICT self-discipline varies depending on the organisation and industry an employee works in. Insights and recommendations for future research are shared in this paper. Findings from this study contribute to IS research and practice.

## 1   Introduction

ICT's are being adopted worldwide at unprecedented rates.  The number of Internet users per 100 inhabitants around the world tripled in the last ten years (ITU 2015). It is expected that by the end of 2015, 96.8% of the world's population will have a mobile device subscription (ITU 2015). Facebook users globally have increased by 200 million over the last year to just shy of a billion daily users (Statista 2015). Without a doubt, technology is changing the way we work, live and communicate (Leung, 2011).

Information and Communication Technologies (ICTs) include the networks, devices and applications used to connect to multiple sources of information to share knowledge (Rice and Leonardi 2013). The "embedded" presence of ICTs has become a common trend across a variety of workplaces (Karr-Wisniewski and Lu 2010; Richardson and Benbunan-Fich 2011). It is becoming difficult to find a work setting in developed countries where employees do not require the support of ICTs to perform their jobs, even unskilled labour tasks often require the use of a mobile phone or tablet device.

The challenge that faces individuals is how to manage their degrees of ICT connectivity to avoid disruptions in individual workflow, and ensure productivity. As Dery et al. (2014) put it

*"understanding [ICT] connective choices in practice represents a significant challenge, but it is certainly a frontier worthy of continued exploration"* (pg. 569).

The study described in this paper aims to assess the influence of ICT self-discipline on the effect of ICT connectivity on employee productivity. Effective personal management practices (stemming from ICT self-discipline) can lead to an improvement in employee productivity and organisational performance. The significance of our study was motivated and reinforced by calls for future research on this topic made in IS, Communications academic literature and in the popular media (Burger 2014; Cruickshank 2014; Derks and Bakker 2010; Leung 2011; Majchrzak et al. 2014; Mazmanian et al. 2006; Wajcman et al. 2010).

This paper is presented as follows: a brief summary of the relevant literature to highlight the key concepts involved, an outline of the research design, our findings and the limitations of the results, wrapped up with a forward looking discussion and conclusion.

## 2    Literature Review

Technology's rapid evolution has introduced a daunting array of communication devices (ICTs) and modes of communication. Collectively, ICT's have brought employees and other sources of information much closer than ever before, expanding employees' ICT connectivity and widening the potential to be reached (Dery and MacCormick 2012; Harmon and Mazmanian 2013; Stephens 2012). In particular, Kolb (2008) explains organizational connectivity as the *"coordination between employees within various types of structural (e.g. centralized–decentralized) and spatial configurations (e.g. traditional offices, telework, hot-desking and virtual organizations) through ICTs"* (pg. 135).

Cecez-Kecmanovic et al. (2014) conceptualise four modes of connectivity that professionals encounter in the workplace:

i. Connectivity that is enabling and inevitable is referred to as 'connected as a form of life'.

ii. Connectivity that is disturbing and inevitable puts professionals in a mode referred to as 'burnt by connectivity'.

iii. Connectivity that is enabling and controllable is referred to as 'restricting connectivity and protecting oneself'.

iv. Connectivity that is disturbing and controllable puts professionals in a mode that makes them 'struggling with connectivity'.

The focus of this research is to move employees from a state of 'struggling with connectivity' to 'restricting connectivity and protecting oneself' by building awareness on the notion of ICT self-discipline to filter the 'good' connectivity from the 'bad' connectivity.

The optimistic discourse on ICT connectivity advocates that it enhances employee productivity by adding speed and flexibility (Coker 2011; Mazmanian 2013; Skeels and Grudin 2009). On the other hand, a more negative view on ICT connectivity will defend that the constant state of connectedness (creating continuous disruptions to workflow, multitasking and technostress) can have adverse effects on employee productivity (Ayyagari et al. 2011; Cameron and Webster 2013; Perlow and Porter 2009; Tarafdar et al. 2013). It has been suggested that employees should aim to find the optimal level of ICT connectivity, that is, the right amount of ICT connectivity in order to fulfill their work needs (Cecez-Kecmanovic et al. 2014; Dery et al. 2014). We believe that the concept of ICT self-discipline underpins this notion.

There is limited research on the notion of ICT self-discipline, as it is an emerging phenomenon (Al-Dabbagh et al. 2014). Therefore, this section provides foundational insights on ICT self-discipline first from Psychology and then from an IS perspective.

In Psychology, the notion of self-control emerged to reflect an individual's desire for immediate gratification (Freud 1911, 1959). Theories such as the General Theory of Crime (Gottfredson and Hirschi 1990) and Self-Regulation Theory (Baumeister and Vohs 2007) posit that high self-control can lead to positive outcomes. For instance, individuals with high self-control are better at regulating impulsive behaviours when consuming alcohol, binge-eating and preparing for school examination (Buker 2011; Tangney et al. 2004). Buker (2011) refers to this as *"an individual's decision or ability to delay immediate gratification of desires in order to reach larger alternative goals"* (pg. 266). In this paper, ICT self-discipline stems from this notion of self-control as it entails a similar phenomenon of regulating behaviours, in this case towards ICTs (Al-Dabbagh et al. 2014).

Based on discussions during the qualitative phase of this study (which is explained in section 3), the term *control* was viewed as a form of power or authority, whereas *discipline* was seen as a notion concerned with choices an individual made. Thus, it was decided to use the term ICT self-*discipline*, since this study focused on employees' own abilities to regulate behaviours.

From an IS perspective, ICT self-discipline is illustrated by the strategies employees undertake at work to ensure ICT connectivity provides a positive effect on their productivity. In response to the workplace shift caused by ICT connectivity, employees are pressured to develop strategies to manage their ICTs as well as information flows (Cameron and Webster 2013). For instance, employees should respond to emails only at certain times of the day or switch their smartphones off after leaving the office (Mazmanian et al. 2013; Perlow 2012). These behaviours may vary depending on the work environment an employee comes from, such as an employee's organizational culture, structure and portfolio of tasks. Further, technology applications such as *Selfcontrol$^{tm}$*, *CanFocus$^{tm}$* and *Asana$^{tm}$* are also used to moderate ICT connectivity in the workplace. Organisations around the world have implemented new workplace policies in an attempt to address the issue of an over-connected workforce (Fishwick 2014; Mangan 2014; Vasagar 2013). These applications and work policies attempt to counteract the negative consequences of ICT connectivity in order to assist increasing productivity.

A stream of research has continuously promoted investigating the notion of self/personal control as a way to develop new strategies to maximize the benefits of ICT connectivity in the workplace (Leung 2011; Derks and Bakker 2010; Rennecker and Godwin 2005; Wajcman et al. 2010). As a result, we took the step to explore the influence of ICT self-discipline in the workplace in order to understand how employees can enhance their performance and well being through the development of effective personal management strategies related to ICT use.

While theories such as Media Richness Theory (Daft and Lengel 1986 - matching mode of communication with equivocality of message), Task-Technology Fit (Goodhue 1995 - matching task with technology capabilities) and Media Synchronicity Theory (Dennis et al. 2008 - matching the required synchronicity of the communication process with the medium's synchronicity capabilities) are useful for understanding the behaviour of individuals, they seem to lack sufficient detail about ICT self-discipline as a concept underlying employee behaviour towards individual ICT connectivity. In response to this gap, the research model (Figure 1) was developed to guide this study.

In this study, ICT connectivity was defined as the extent to which an individual is connected to ICTs (Dery et al. 2014; Leung 2011). This does not include emotional connectedness, rather, it considers the employer's physical connection to applications, devices and networks, regardless of whom they have been provided by. Employee productivity was defined as the extent to which an individual perceives him/herself to have accomplished the expected work during a typical work-day (Torkzadeh and Doll 1999). ICT self-discipline was defined as an individual's ability to regulate their behaviours towards ICTs (Al-Dabbagh et al. 2014). The model suggests hypothesis 1 (H1): ICT connectivity influences employee productivity. This is suggested in research highlighting the mixed effects of ICTs that employees experience (Ayyagari et al. 2011; Mazmanian 2013). The model also suggests hypothesis 2 (H2): ICT self-discipline influences the effect of ICT connectivity on employee productivity. This is suggested by ongoing concern in the literature on the influence of the management of information flow through ICTs (Leung 2011; Derks and Bakker 2010; Rennecker and Godwin 2005; Wajcman et al. 2010).

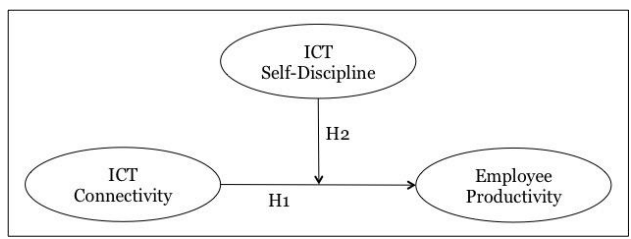

*Figure 1: Research Model*

The following section explains the research design.

## 3  Research Design

Initially, interviews and focus groups took place to refine the concept of ICT self-discipline and to develop the items for the online survey (Krueger and Casey 2009; Myers and Newman 2007). Further details on the qualitative phase of this study are presented in (Al-Dabbagh et al. 2014). Relevant literature was assessed to further develop the survey items (Tangney et al. 2004; Torkzadeh and Doll 1999; Yun et al. 2012). The items were then tested and validated through a pilot survey and expert reviews (Straub et al. 2004). The survey items were refined before the online survey link was distributed to participants.

An online survey was developed and used to test the influence of ICT self-discipline on the effect between ICT connectivity and employee productivity (Evans and Mathur 2005). The research population consisted of selected employees from New Zealand organisations whose roles involved carrying out ICT-supported business tasks (Gebauer et al. 2007). Individuals from a wide range of jobs and working environments were approached to ensure variety (Becker 1998).

The data was analysed through statistical testing using SPSS software to verify the measures of the research phenomena (Gefen and Straub 2005; Hinkin 1998; Straub 1989). In addition, SmartPLS 3.0 software was used to conduct a factor analysis and the assessment of the reliability of the measurement model using Cronbach's alpha (Hinkin 1998). Specifically, variance-based Structural Equation Modeling Partial Least Squares (SEM-PLS) was used to assess the relationships amongst the phenomena (Hair et al. 2011). This technique was used as the capabilities of the method best aligned with the nature of the research (Ringle et al. 2012). The following section shares the survey findings.

## 4  Findings

The online survey (see Appendix 1) was made available for eight weeks and yielded 443 useable responses. The online survey sample consisted of 35% females and 65% male. Respondents came from industries including Information technology (IT), education, business and consulting, architecture, law and finance. Seventy-three percent of the participants were full-time employees, coming from either a small or large organisation.

Construct reliability was tested for to seek the degree to which each construct measured what it was supposed to (Venkatesh et al. 2013). The Cronbach's alpha was calculated (target >0.7). All the research constructs exceeded thresholds for the Cronbach's alpha test, as presented in Appendix 1.

Indicator reliability was tested for to seek the degree the items within a construct correlated. The thresholds recommended in literature were met - e.g. factor loadings >0.6 and cross-loadings <0.4 (Nunnally and Bernstein 1994; Huber et al. 2007). In addition, all items loaded into their appropriate construct and exceeded the >0.6 threshold, apart from one item (IP05), which was kept due to its marginal value and relevance (Nunnally and Bernstein 1994). The results are presented below (cross-loadings are not shown as none exceeded the specified threshold).

| Item | ICT Connectivity | ICT Self-Discipline | Employee Productivity |
|---|---|---|---|
| ICTC01 | 0.872 | | |
| ICTC02 | 0.723 | | |
| ICTC04 | 0.819 | | |
| ICTC06 | 0.881 | | |
| IP01 | | | 0.746 |
| IP02 | | | 0.779 |
| IP03 | | | 0.813 |
| IP04 | | | 0.729 |
| IP05 | | | 0.587 |
| SD01 | | 0.722 | |
| SD02 | | 0.638 | |
| SD06 | | 0.702 | |
| SD08 | | 0.757 | |
| SD09 | | 0.614 | |
| SD10 | | 0.827 | |

*Table 1.  Factor Loadings and Cross Loadings*

The research hypotheses were tested through SEM-PLS (Chin and Newsted, 1999). The path coefficients and significance (t-stat) were tested for. Hypothesis H1 was supported (path = 0.087, t-stat = 1.818 *p < 0.1*). A product indicator moderation test was conducted to assess the influence of ICT self-discipline in the research model (H2). This type of moderation test was conducted as it fit best with the size of research sample (Hair et al. 2013; Henseler and Fassott 2010). The influence of ICT self-discipline was not significant on the entire research sample (Table 2), but the size of the path coefficient and graphical representation (Figure 2) showed that ICT self-discipline had a reasonable influence on the effect between ICT connectivity and employee productivity.

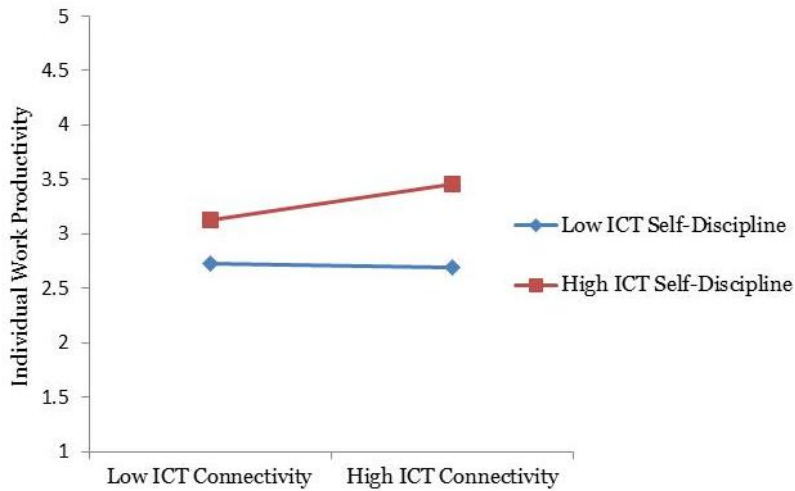

*Figure 2: Moderating Effect of ICT Self-Discipline (Entire dataset)*

Figure 2 clearly illustrates the influence of ICT self-discipline in the workplace. The red line shows that when ICT connectivity increases in a work environment then employee productivity should increase, *if* a strict (or high) level of ICT self-discipline is exercised. On the other hand, the blue line in Figure 2 suggests that when ICT connectivity increases in a work environment then employee productivity can decrease, *if* a lenient (or low) level of ICT self-discipline is exercised.

Just because a hypothesis is not supported it does not mean one should ignore an effect in a model as a whole. The regression weights are not significant enough to say that the 'null' hypothesis is not true (Field 2013). Therefore, in order to provide further clarification on H2, the hypothesis was tested on sub-groups of the survey sample. The two top industry segments were IT ($n = 151$) and education ($n = 71$). Given the differences in the nature of their work roles it was decided to assess the influence of ICT self-discipline on each of the industries separately. Additionally, employees from large organisations ($n = 52$) and small organisations ($n = 61$) were assessed independently given the differences in the organisational structure. Other sample sub-groups were not assessed due to having an insufficient sample size for SEM-PLS analysis, that is, less than 50 datasets (Chin and Newsted 1999). The test results are presented in Table 2. The results are explained in the discussion of this paper. The tests were also run for different gender and age groups, however the statistical results yielded no significance.

| Group | Path Coefficient | T-Stat |
|---|---|---|
| Entire dataset | 0.119 | 0.628 |
| IT industry | -0.265 | 1.007 |
| Education industry | 0.620 | 2.972** |
| Small organisation | -0.848 | 1.132 |
| Large organisation | 0.726 | 0.954 |

*\*\*p < 0.01.*

*Table 2. Testing the Influence of ICT Self-Discipline (H2)*

Further, an Impact Performance Matrix Analysis (IPMA) was conducted through SmartPLS 3.0 to identify 'areas' for improvement that should be addressed by management in organisations (Ahmad and Afthanorhan 2014; Hair et al. 2013; Hock et al. 2010). The results from the IPMA indicated that ICT self-discipline had high impact on employee productivity, but low performance in the test. This means that ICT self-discipline needs to be 'enhanced' in order to increase employee productivity. The following section discusses the research findings.

# 5 Discussion

The purpose of this study was to assess the influence of the emerging notion of ICT self-discipline on the effect of ICT connectivity on employee productivity.

The observed effect between ICT connectivity and employee productivity in this study would be generally classified 'weak' in IS research (Cohen, 1988). This weak effect could be explained as the cancelling effect of the collective positive and negative consequences of ICTs (mentioned in the literature review). From a holistic perspective, the positive consequences can be seen as intuitively outweighing the negative consequences and interpreted as an overall positive outcome. This insight sheds a generally positive light on the effect of ICT connectivity on employee productivity, whereas previous studies have painted a negative picture of this effect (Ayyagari et al. 2011; Hung et al. 2011; Tarafdar et al. 2007).

Statistical tests were conducted to investigate the influence of ICT self-discipline on the effect between ICT connectivity and employee productivity. From a high-level perspective, ICT self-discipline has a positive moderating influence on the effect between ICT connectivity and employee productivity. This means that as ICT self-discipline becomes stricter (increases), the effect of ICT connectivity on employee productivity improves (increases). The significance of this relationship was too minimal to support the hypothesis either concretely or holistically. This may have been due to the survey sample covering such a wide range of industries, making it challenging to generalise the influence of ICT self-discipline without the occupational contextualization described above.

## 5.1 Testing H2 for different industries

Hypothesis H2 yielded different results when it was tested in sub-samples of the research population. In the sample of employees only from the education industry, ICT self-discipline placed a positive and relatively significant influence on the effect of ICT connectivity on employee productivity.

Employees from the education industry sample included lecturers, librarians, teachers, trainers and research assistants. The job descriptions of these roles had low interdependence confirmed via a New Zealand government website which provided definitions of common industry roles (www.careers.govt.nz). When these descriptions were assessed, it appeared that the nature of these roles had low- medium interdependence. This meant that employees required little input from others to perform their job (Gebauer et al. 2010). Particularly, this included employees with relatively autonomous jobs (such as teaching) that could be performed without frequent coordination and interaction with others, and pooled jobs (such as researching) that required two or more people to interact to perform the job (Hackathorn and Keen 1981).

Thus, having a somewhat independent nature of work is expected to give employees the freedom to ignore communication through ICTs from time to time to avoid interruptions without the risk of missing out on relevant information, hence improving productivity. This explained why ICT self-discipline was a positive moderator on the effect between ICT connectivity and employee productivity.

On the other hand, in the sample of employees from the IT industry, ICT self-discipline placed a negative influence on the effect of ICT connectivity on employee productivity. The influence was not significant in this context. While in this case H2 was not supported the results still provided meaningful insight. The roles in the IT industry sample included software developers, project managers, business analysts, software testers and consultants. These jobs were defined via www.careers.govt.nz. The nature of these roles had medium-high interdependence, which meant employees required input from others in order to perform their jobs. To be exact, this included employees with pooled jobs (such as software testing) that required two or more people to interact to perform the job, and sequential jobs (such as project managers) where employees can only perform their jobs with a sequence of inputs coming from other sources of information (Hackathorn and Keen 1981).

To illustrate the discussion above, a project manager may not be able to impose strict ICT self-discipline as that might lead to a loss of critical information necessary for the job. ICT self-discipline may enhance an employee's productivity at one point in time but it may also increase the backlog of

emails, which in the long run could result in a negative effect on productivity. This explained why ICT self-discipline was a negative moderator on the effect between ICT connectivity and employee productivity.

## 5.2 Testing H2 for different organisation structures

There were similar contrasting differences between the results of H2 when assessed for employees from large organisations and for employees from small organisations. Employees from a large organisation faced a positive and medium-strong influence from ICT self- discipline on the relationship between ICT connectivity and employee productivity. Contrastingly, employees from small organisations faced a negative influence from ICT self-discipline on the relationship between ICT connectivity and employee productivity.

The contrasting results mentioned above are likely to be due to the nature of the organisation the employees worked in. Large organisations are seen to have formal structures, whereas small organisations are seen to have informal ones (Chen and Hambrick 1995; Russo and Perrini 2010).

Further, having a process-driven and strictly regulated environment in a large organisation makes it suitable for employees to up-take strict ICT self- discipline, hence the positive moderation of H2. On the other hand, small organisations are seen to have a more flexible work environment, which would seem less inviting for strict ICT self-discipline. This explained the negative moderation of H2.

Additionally, large organisations are likely to have more resources (McAdam and Reid 2001) and face more communication through ICT connectivity compared to small organisations. This implied that ICT self- discipline would assist in filtering out the 'good' and the 'bad' connectivity in such inter-connected work environments, explaining the positive moderation of H2 for large organisations.

Although the results of ICT self-discipline in large and small organisations were insignificant to support H2, they still provided useful suggestions on the differences in the role of ICT self-discipline in these work contexts.

## 5.3 Overall views on ICT self-discipline

The findings from this study emphasize that the effect of ICT self-discipline in the workplace is subject to the work context (job type and organisation type) an employee is in. This finding was not previously indicated in the literature on individual/self control. Contributing specific work factors to assess will help employees tailor ICT self-discipline to their needs so they obtain the best from ICT connectivity and receive added productivity at work.

Literature on IS suggests that external factors beyond an individual's control, such as work conditions, can influence the way employees managed ICTs (Anandarajan et al. 2000; Loges and Jung 2001; Markus 1983; Orlikowski 2008; Ou and Davison 2011; Rossi 2002). Thus, it made sense that an employee's job and organisational structure can influence their level of ICT self- discipline.

The findings from this study also indicated the importance of ICT self-discipline, particularly by the IPMA and the effect size test in the sub-group analysis. ICT self-discipline is a positive and significant moderator of the relationship between ICT connectivity and employee productivity in job types that have low-medium interdependence (such as jobs in the education industry). Further, ICT self-discipline can be seen as a positive moderator in organisations with low flexibility and that have strict regulations in place (such as large organisations).

On the other hand, ICT self-discipline can be seen as a negative moderator of the relationship between ICT connectivity and employee productivity in work industries that have high interdependence (such as the IT industry) and in organisations with high flexibility (such as small organisations).

## 5.4 Recommendations for practitioners

Employees need to find the optimal ICT connectivity, that is, the right amount of ICT connectivity in order to fulfill their work needs (Cecez-Kecmanovic et al. 2014; Dery et al. 2014). We believe that employees can only work towards this optimal ICT connectivity by exerting the appropriate levels of ICT self-discipline at work. Employees should focus on applying disciplined behaviours based on their work environments and requirements to assist them in getting rid of the 'bad' connectivity and taking full advantage of the 'good' connectivity. It is a matter of filtering information through ICT connectivity and finding the right balance. To help find the right balance, the following strategies are recommended for practitioners:

- prioritise and carefully manage incoming-exchanges (while taking advantage of tool notifications) to avoid interruptions to workflow and maintain productivity,
- make quick and ruthless decisions with how responsive to be through ICTs (i.e. a response likely to take longer than two minutes should be deferred),
- have a filing strategy for tasks (i.e. group based on importance),
- pay less attention to less important emails (i.e. the importance of an email is higher if an employee is in the "to" addressee list compared to an employee in the "cc" addressee list),
- be mindful of colleagues' productivity and avoid interrupting others where possible,
- align the task at hand with the appropriateness of the mode of communication, particularly if there is urgency in the task then a more synchronous mode of communication should take place (i.e. a phone call or face-to-face communication) and less urgent tasks should be sent via email).

### 5.5 Future research

The findings from this research address the ongoing concern in the literature and in the media about the mixed effects of ICT connectivity in the workplace. The research highlights the importance of ICT self-discipline in the workplace and how it can enhance the effects of ICTs on employees. Further, the findings suggest critical factors that organisations should assess prior to enforcing or suggesting ICT self- discipline strategies for work.

The results of this study suggest new avenues for future research, particularly focusing on factors like job types, work settings, expectations and personality types and their effect on the research phenomena. Future research is needed to explore the concept of ICT self-discipline and tailor the contributed research model to varying contexts. We suggest the incorporation of theories such as the Theory of Reasoned Action to explain the behaviours towards discipline (Ajzen and Fishbein 1980). Self-Regulation Theory (Baumeister and Vohs 2007) could also provide reasons behind the behaviours of ICT self-discipline, explaining the short-term desires of impulsive behaviours. Future research should also investigate the topic through other approaches such as in-depth qualitative research or longitudinal studies to gain further understanding.

Managers are encouraged to be aware of employees' behaviours towards ICTs and should train them on how to face more positive experiences with ICTs. It is critical that the employee's job characteristics and organisational structure are assessed prior to imposing workplace strategies on how to manage ICT connectivity at work.

## 6 Conclusion

The primary goal of this research was to investigate the influence of ICT self-discipline on the effect between ICT connectivity and employee productivity. The survey findings provided novel insights for academia and professional practice, highlighting the influence and importance ICT self-discipline has for different work contexts.

The study contributes a new model towards the understanding of the use of ICTs in the workplace. The model acts as a basis and can be expanded to incorporate other relevant factors and foundational theories. The research also expands knowledge on the emerging phenomenon of ICT self-discipline, sometimes described as self-control or individual control. This research also answers prior calls for more research made by academics and in public media to explain the impact ICTs are having on employees. These insights, along with a reliable survey instrument for capturing the phenomena investigated in this research, are contributed to the literature on IS, Communications, Organisational Studies and Psychology.

From a practitioner viewpoint, this study contributes the awareness of a further developed phenomenon, ICT self-discipline, which is relevant for enhancing employee productivity within the ICT-connected workplace. Our research results also provide a theoretical basis for developers interested in building applications that can help manage ICT connectivity to enhance employee productivity.

It is still possible that errors may have taken place during the quantitative measurement of this research. To overcome these limitations, future research is encouraged to repeat similar research from a different theoretical outlook or via other research methods. Future research can also investigate this topic through richer data-gathering approaches to capture any influential factors that may have been missed. This study was conducted at a one point in time and limited to a New Zealand context. Thus, further investigation on this phenomenon could explore other work contexts and be carried out longitudinally to capture the changing of behaviours and attitudes towards ICTs. We strongly encourage further development of this crucial phenomenon of ICT self-discipline and its impact on the workplace and other contexts.

# Appendix 1

| Construct | Item |
| --- | --- |
| ICT connectivity<br><br>Cronbach's alpha = 0.849 | I am connected to ICTs the entire time. |
| | I always have my ICTs with me everywhere I go. |
| | I perceive myself to have a high level of ICT connectivity. |
| | I spend all of my day connected through ICTs. |
| Employee productivity<br><br>Cronbach's alpha = 0.788 | I always accomplish the work that I expected to. |
| | I always accomplish my work within the time allocated for it. |
| | I am always productive. |
| | I always accomplish more work than I had expected. |
| | I hardly ever get my work done on time.* |
| ICT self-discipline<br><br>Cronbach's alpha = 0.816 | I am good at ignoring incoming communication through ICTs, even if I am tempted to check them. |
| | It is hard to stop myself from using ICTs even if I know they are unnecessary.* |
| | I am always able to refuse communications through ICTs that are not immediately relevant for my work. |
| | I am highly disciplined when using my ICTs. |
| | I find it very difficult to ignore my ICTs when they are nearby.* |
| | I am always able to stay focused and do not let ICTs interrupt me. |

*Note.* * = reverse-coded item.

# Copyright

The following copyright paragraph must be appended to the paper. Author names should not be included until after reviewing. Please ensure the hyperlink remains for electronic harvesting of copyright restrictions.